\documentclass[%
 aps,
 prl,
 twocolumn,
 reprint,
 amsmath,amssymb,
 superscriptaddress,
]{revtex4-2}

\usepackage{graphicx}
\usepackage{braket} 

\usepackage{float}
\usepackage{dcolumn}
\usepackage{bm}

\usepackage[mathlines]{lineno}
\usepackage{hyperref}
\usepackage[dvipsnames]{xcolor}

\begin{document}


\title[EIT cooling]{Electromagnetically-Induced-Transparency Cooling of High-Nuclear-Spin Ions}
\author{Chuanxin Huang}
\thanks{These authors contribute equally to this work}%
\affiliation{Center for Quantum Information, Institute for Interdisciplinary Information Sciences, Tsinghua University, Beijing 100084, P. R. China}
\author{Chenxi Wang}
\thanks{These authors contribute equally to this work}%
\affiliation{Center for Quantum Information, Institute for Interdisciplinary Information Sciences, Tsinghua University, Beijing 100084, P. R. China}
\author{Hongxuan Zhang}
\affiliation{Center for Quantum Information, Institute for Interdisciplinary Information Sciences, Tsinghua University, Beijing 100084, P. R. China}
\author{Hongyuan Hu}
\affiliation{Center for Quantum Information, Institute for Interdisciplinary Information Sciences, Tsinghua University, Beijing 100084, P. R. China}
\author{Zuqing Wang}
\affiliation{Center for Quantum Information, Institute for Interdisciplinary Information Sciences, Tsinghua University, Beijing 100084, P. R. China}
\author{Zhichao Mao}
\affiliation{HYQ Co., Ltd., Beijing 100176, P. R. China}
\author{Shijiao Li}
\affiliation{Center for Quantum Information, Institute for Interdisciplinary Information Sciences, Tsinghua University, Beijing 100084, P. R. China}
\author{Panyu Hou}
\affiliation{Center for Quantum Information, Institute for Interdisciplinary Information Sciences, Tsinghua University, Beijing 100084, P. R. China}
\affiliation{Hefei National Laboratory, Hefei 230088, P. R. China}
\author{Yukai Wu}
\affiliation{Center for Quantum Information, Institute for Interdisciplinary Information Sciences, Tsinghua University, Beijing 100084, P. R. China}
\affiliation{Hefei National Laboratory, Hefei 230088, P. R. China}
\author{Zichao Zhou}
\email{zichaozhou@mail.tsinghua.edu.cn}
\affiliation{Center for Quantum Information, Institute for Interdisciplinary Information Sciences, Tsinghua University, Beijing 100084, P. R. China}
\affiliation{Hefei National Laboratory, Hefei 230088, P. R. China}
\author{Luming Duan}
\email{lmduan@tsinghua.edu.cn}
\affiliation{Center for Quantum Information, Institute for Interdisciplinary Information Sciences, Tsinghua University, Beijing 100084, P. R. China}
\affiliation{Hefei National Laboratory, Hefei 230088, P. R. China}
\affiliation{New Cornerstone Science Laboratory, Beijing 100084, PR China}
\date{\today}

\begin{abstract}
We report the electromagnetically-induced-transparency (EIT) cooling of $^{137}\mathrm{Ba}^{+}$ ions with a nuclear spin of $I=3/2$, which are a good candidate of qubits for future large-scale trapped ion quantum computing.
EIT cooling of atoms or ions with a complex ground-state level structure is challenging due to the lack of an isolated $\Lambda$ system, as the population can escape from the $\Lambda$ system to reduce the cooling efficiency.
We overcome this issue by leveraging an EIT pumping laser to repopulate the cooling subspace, ensuring continuous and effective EIT cooling. We cool the two radial modes of a single $^{137}\mathrm{Ba}^{+}$ ion to average motional occupations of 0.08(5) and 0.15(7) respectively. Using the same laser parameters, we also cool all the ten radial modes of a five-ion chain to near their ground states.
Our approach can be adapted to atomic species possessing similar level structures. It allows engineering of the EIT Fano-like spectrum, which can be useful for simultaneous cooling of modes across a wide frequency range, aiding in large-scale trapped-ion quantum information processing.

\end{abstract}

\maketitle

Trapped ions are a powerful and versatile platform, enabling a broad range of research in quantum information processing (QIP)~\cite{10.1063/1.5088164,RevModPhys.93.025001}, precision spectroscopy~\cite{RevModPhys.87.637}, and tests of fundamental physics~\cite{RevModPhys.90.025008}. Ions can be controlled by laser and microwave fields, and can achieve high-fidelity operations including state preparation and measurement~\cite{harty2014high,christensen2020high,ransford2021weak,erickson2021high,an2022high,Moses2023}, single-qubit gates~\cite{harty2014high,brown2011single}, and two-qubit gates~\cite{PhysRevLett.117.060504,PhysRevLett.117.060505,srinivas2021high,clark2021high}, for up to tens of ions \cite{Egan2021,Postler2022,Moses2023,chen2023benchmarking}.
Hundreds of ions have also been used for quantum simulation in Paul traps \cite{guo2023site} and Penning traps \cite{britton2012engineered,bohnet2016quantum}.

For laser-based gate operations, thermal motion of ions causes random fluctuation in laser phase and intensity, fundamentally limiting gate fidelity. Although advanced techniques such as the widely used Molmer-Sorensen entangling gate \cite{PhysRevA.62.022311,Milburn2000MSgate} alleviate the requirement of ground state cooling, it is still desirable to cool the relevant motional modes of ions to the lowest possible temperature for achieving high fidelity QIP.
Among the common sub-Doppler cooling methods, resolved sideband cooling has been routinely applied to small ion crystals, typically achieving an average motional qunatum (phonon) number $\bar{n}<0.1$~\cite{wineland1998experimental,RevModPhys.75.281}. However, due to the narrow cooling band, its cooling time generally grows with the number of ions~\cite{PhysRevA.102.043110,PhysRevResearch.5.023022}, which is inefficient for large ion crystals.
Polarization gradient cooling~\cite{Birkl_1994,PhysRevLett.119.043001,Joshi_2020} offers a broader cooling band but the attainable $\bar{n}\approx 1$ is high~\cite{Dalibard:89}.
In contrast, electromagnetically-induced-transparency (EIT) cooling \cite{RevModPhys.75.281,PhysRevLett.85.4458} provides high cooling rates and broad cooling bandwidth, while maintaining a low cooling limit $\bar{n}\approx 0.1$, making it a suitable cooling method for quantum computing \cite{PhysRevLett.97.050505} and quantum simulation \cite{RevModPhys.93.025001} with large ion crystals.

EIT cooling has been demonstrated with ions that have a relatively simple level structure including $^{24}\mathrm{Mg}^{+}$~\cite{PhysRevLett.110.153002} and $^{40}\mathrm{Ca}^{+}$~\cite{PhysRevLett.85.5547,PhysRevApplied.18.014022} with no nuclear spins, and $^{171}\mathrm{Yb}^{+}$ with a nuclear spin $I=1/2$~\cite{PhysRevLett.125.053001,PhysRevLett.126.023604}. It has also been realized with $^{9}\mathrm{Be}^{+}$ possessing $I=3/2$ but under a strong magnetic field (4.46\,T)~\cite{PhysRevLett.122.053603}, such that only electron spin states are relevant.
However, for ions with high nuclear spins $I>1/2$ at a low or intermediate magnetic field, EIT cooling is challenging due to the difficulty of finding an isolated $\Lambda$ level structure. Despite this difficulty, many high-nuclear-spin ion species such as $^{9}\mathrm{Be}^{+}$, $^{25}\mathrm{Mg}^{+}$, and $^{43}\mathrm{Ca}^{+}$ have been widely used for QIP due to their long coherence times and high-fidelity quantum operations~\cite{an2022high,PhysRevResearch.2.033128,PhysRevLett.117.060504,PhysRevLett.117.060505,PhysRevA.77.062306,Tan2015,PhysRevA.81.052328}. In particular, $^{137}\mathrm{Ba}^{+}$ ions with $I=3/2$ recently have attracted increasing interest due to the required laser transitions in the visible wavelength range~\cite{an2022high,PhysRevA.81.052328,PhysRevResearch.2.033128}. This provides key benefits for large-scale QIP, as high laser power is more accessible at this range and these lasers cause less photo-induced charging~\cite{wang2011laser} and damage in optics compared to lasers near or within ultraviolet range used for other ion species mentioned above.

\begin{figure*}
\includegraphics[width=0.95\textwidth]{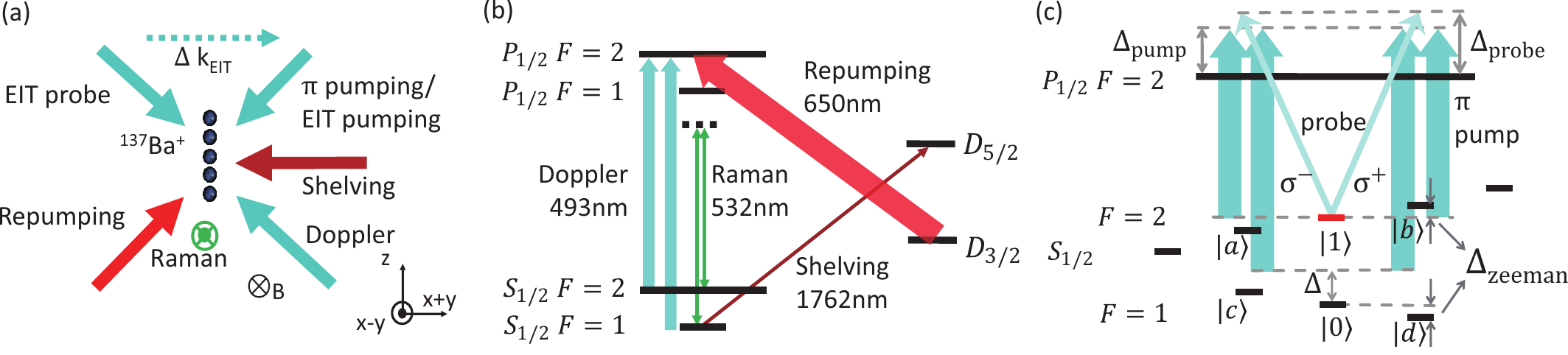}
\caption{
(a) Experiment configuration and (b) relevant energy levels of $^{137}\mathrm{Ba}^{+}$ ions.
We use $493\,$nm lasers for Doppler and EIT cooling.
A $650\,$nm laser repumps the population in the $D_{3/2}$ level to the $P_{1/2}$ level. A $1762\,$nm laser transfers the population in $|0\rangle$ [see (c)] to $D_{5/2}$ for state readout. A pair of focused $532\,$nm Raman laser beams are use to drive sideband transitions for measuring average phonon numbers.
(c) Relevant energy levels and laser couplings of the EIT cooling scheme. A weak probe beam (narrow cyan arrows) has $\sigma_{+}$ and $\sigma_{-}$ polarization components. A strong $\pi$-polarized pump beam (broad cyan arrows) consists of two frequency components separated by approximately the hyperfine splitting of $8.037\,$GHz. In the limit of weak probe, the ion mostly stays in the $\ket{1}$ state with EIT cooling occurring between this state and the four nearby states $\ket{a}$, $\ket{b}$, $\ket{c}$ and $\ket{d}$.
}
\label{fig:trap}
\end{figure*}

In this paper, we demonstrate EIT cooling with $^{137}$Ba$^{+}$ ions. We modify a laser configuration used for qubit state initialization to perform EIT cooling in an open $\Lambda$ system. Despite all eight states in the ground-state manifold that are involved in the dynamics during EIT cooling, EIT pump beams, which also serve as repumping beams, keep population in the cooling subspace. (See Supplemental Material Sec.~V \cite{supplemental} for generalization to other high-nuclear-spin ion species.)
We cool two radial modes of a single $^{137}$Ba$^{+}$ ion to  $\bar{n}$ of 0.08(5) and 0.15(7), respectively.
By using the same laser parameters, we further cool all radial modes of a five-ion chain to near the ground states.

The experimental system and laser couplings are depicted in Fig.~\ref{fig:trap}. We confine up to five $^{137}$Ba$^{+}$ ions in a blade trap with trap frequencies $(\omega_{x},\,\omega_{y},\,\omega_{z}) = 2\pi\times(1.7,\,1.8,\,0.2)\,$MHz.
We denote the ion chain axis as $z$ coinciding with the trap axis.
%
A quantization field of approximately $6.7\,$Gauss is applied perpendicular to the $z$ axis, leading to a Zeeman splitting of $\Delta_{\mathrm{Zeeman}}=2\pi\times 4.7$\,MHz between adjacent $S_{1/2}$ Zeeman levels.
A Doppler cooling beam at 493\,nm has all the three polarization components and couples to modes in all the three principle axes.  It has two frequency components that respectively couple the $S_{1/2}, F=1$ and $S_{1/2}, F=2$ states to $P_{1/2}, F=2$ with a detuning of $-\Gamma/2$, where $\Gamma=2\pi\times 20.1\,$MHz is the natural linewidth of the $P_{1/2}$ states.
A $\pi$-polarized $493\,$nm laser, used for optical pumping [$\pi$ pumping in Fig.~\ref{fig:trap}(a)], also has two frequency components to resonantly couple $S_{1/2}, F=1,2$ states to $P_{1/2}, F=2$, respectively. Owing to the selection rule, the ion will be pumped to the dark state $\ket{1}\equiv \ket{S_{1/2},F=2,m_F=0}$ ideally. This beam also serves as a pump beam in EIT cooling which will be described later.
During the cooling and optical pumping cycles, ions can decay to $D_{3/2}$ and remain dark. We use a strong near-resonant $650\,$nm laser (repumping beam) containing multiple frequency components to deplete the population in all the hyperfine levels in $D_{3/2}$.
A pair of counter-propagating $532\,$nm Raman laser beams can drive Raman sideband transitions between $\ket{0} \equiv \ket{S_{1/2},F=1,m_F=0}$ and $\ket{1}$ for thermometry of ions. Their beam waist radius is approximately $2\,\mu$m, which is smaller than the distance between neighbouring ions in a chain.
To distinguish $\ket{0}$ and $\ket{1}$, we first apply three $1762\,$nm $\pi$ pulses to sequentially shelve the population in $\ket{0}$ to three different Zeeman levels in $D_{5/2}$, which can suppress the shelving error from imperfect $\pi$ pulses \cite{an2022high}. Then, we turn on the Doppler cooling laser with its frequency adjusted to resonance.  Ions only fluoresce when in $\ket{1}$ and fluorescence photons are collected by a photomultiplier tube for a single ion or an electron-multiplying CCDs for multiple ions. The typical state preparation and measurement error is about 2\% dominated by polarization errors in the $\pi$ pumping beam for state initialization.

As shown in Fig.~\ref{fig:trap}(a), the EIT laser beams consist of a strong pump beam and a weak probe beam. Their wave vector difference $\Delta\boldsymbol{k}_{\mathrm{EIT}}$ is perpendicular to the $z$ axis and has roughly equal projections onto the $x$ and $y$ axes.
The $\pi$-polarized EIT pump beam is the same beam for optical pumping with its frequency shifted by $\Delta_{\mathrm{pump}}$ as shown in Fig.~\ref{fig:trap}(c). We denote the Rabi frequencies of its two frequency components by $\Omega_{\mathrm{pump},1(2)}$ according to their coupling to $S_{1/2},F=1(2)$ levels, with $\Omega_{\mathrm{pump},2}/\Omega_{\mathrm{pump},1}\approx 2$ controlled by an electro-optic modulator. (See Supplemental Material Sec.~III \cite{supplemental} for more details.)
Their relative detuning can be adjusted to fine-tune the Fano-like spectrum for optimizing EIT cooling.
The probe beam, containing equal $\sigma_{+}$ and $\sigma_{-}$ components, couples $\ket{1}$ to the $P_{1/2},F=2$ levels with a detuning $\Delta_{\mathrm{probe}}$ and a Rabi rate $\Omega_{\mathrm{probe}}$. Note that the Rabi frequency used here only represent the transition strength between the $S_{1/2}$ and $P_{1/2}$ levels. For the Rabi frequencies between detailed Zeeman levels, the Clebsch-Gordan coefficients should further be multiplied (see Supplemental Material Sec. I).

\begin{figure}[!tbp]
	\includegraphics[width=0.9\linewidth]{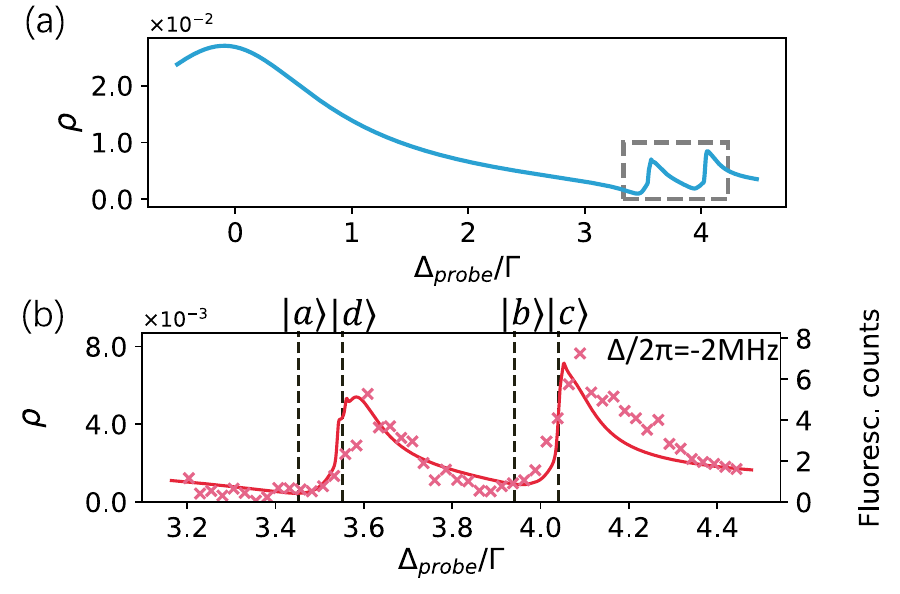}%
	\caption{(a) Fano-like spectrum from numerical simulations when the two frequency components of the EIT pump beam are separated by $\omega_{01}-\Delta$, where $\omega_{01}$ is the qubit transition frequency and  $\Delta/2\pi=-2$\,MHz. The broad peak around $\Delta_{\mathrm{probe}}=0$ corresponds to the resonance of $\ket{1}\rightarrow \ket{P_{1/2},F=2}$. The Fano-like lineshapes associated with EIT in the grey box are zoomed in (b) and compared to the experimental spectrum (crossing marks, after subtracting the background of $7.5$ photon counts).  Vertical dashed lines indicate the dark resonances of four possible $\Lambda$ systems containing $\ket{1}$ and one of \{$\ket{a}$, $\ket{b}$, $\ket{c}$, $\ket{d}$\}.
	}
	\label{fig:Fanospec}
\end{figure}

\emph{EIT cooling scheme---}
In the limit $\Omega_{\mathrm{probe}}\ll \Omega_{\mathrm{pump},1(2)},\Gamma$, the steady state when all EIT beams are on is $\ket{1}$. A weak probe beam can lead to some excitation to the $P_{1/2}$ levels. When the dark resonance condition  $\Delta_{\mathrm{probe}}=\Delta_{\mathrm{pump}}\pm \Delta_{\mathrm{Zeeman}}$ or $\Delta_{\mathrm{probe}}=\Delta_{\mathrm{pump}} + \Delta \pm \Delta_{\mathrm{Zeeman}}$ is satisfied, we expect coherent population trapping in $\ket{1}$ and one of the four nearby states denoted as $\ket{a}$, $\ket{b}$, $\ket{c}$ and $\ket{d}$, so that the excitation to the $P_{1/2}$ level is suppressed.
Fano-like spectra associated with this phenomenon can be observed by tuning the probe beam frequency across the dark resonances, for example the spectrum shown in Fig.~\ref{fig:Fanospec}(a).
Such a Fano-like spectrum can be used to estimate the EIT cooling limit $\bar{n}_f$ and cooling rate $\gamma$ (in the absence of heating from environment), which are given by \cite{RevModPhys.75.281}
\begin{equation}
\bar{n}_f=\frac{\rho(\Delta_{\mathrm{probe}})+\rho(\Delta_{\mathrm{probe}}-\omega)}{\rho(\Delta_{\mathrm{probe}} + \omega)-\rho(\Delta_{\mathrm{probe}}-\omega)},\label{eqa:phonon}
\end{equation}
and
\begin{equation}
\gamma=\eta^{2}\Gamma[\rho(\Delta_{\mathrm{probe}} + \omega)-\rho(\Delta_{\mathrm{probe}}-\omega)],\label{eqa:gamma}
\end{equation}
where $\rho$ is the population of the $P_{1/2}$ states, $\omega$ is the phonon mode frequency, and $\eta$ is the Lamb-Dicke parameter of the EIT beams with a wave vector difference $\boldsymbol{\Delta k}_{\mathrm{EIT}}$.

The key difference from ions with $I=0$ or $1/2$ is that high-nuclear-spin ions have a number of extra ground states and excited states to which ions can escape from the EIT cooling subspace, reducing or diminishing cooling efficacy (See Supplemental Material Sec.~IV \cite{supplemental} for more details). In previous works with $^{171}$Yb$^{+}$ ($I=1/2$)~\cite{PhysRevLett.125.053001,PhysRevLett.126.023604}, the leakage issue was addressed by employing an additional repump laser with the desired ground states protected by the selection rule.
However, for $^{137}\mathrm{Ba}^{+}$, adding repumpers can potentially destroy the ground states used in EIT cooling through their coupling to the extra Zeeman levels of $P_{1/2}$.
Here, we design an EIT cooling scheme where the EIT pump laser simultaneously serves as the repump beam to ensure the cooling efficiency.

In Fig.~\ref{fig:Fanospec} we compute the Fano-like spectrum to guide the design of cooling parameters and compare it with experimental results.
In the simulation, we consider all hyperfine states in the $S_{1/2},F=1,2$ and $P_{1/2},F=2$ manifolds. The $D_{3/2}$ states are ignored since they are efficiently depleted, and we treat this effect as a reduction of the $P_{1/2}$ spontaneous emission rate $\Gamma=2\pi\times 20.1\,$MHz by multiplying the branching ratio of $0.732$ from $P_{1/2}$ to $S_{1/2}$ states \cite{PhysRevA.100.032503}. In general, the interplay of multiple energy levels with multiple laser frequency components is solved by a Lindblad master equation with a time-dependent Hamiltonian. Here we simplify the calculation by computing the spectra for the $\sigma^+$ and $\sigma^-$ components of the probe beam separately and add them up together. Then, each spectrum can be obtained efficiently by solving a time-independent Lindblad master equation after going into a suitable rotating frame. This approximation is valid to the first order for a weak probe beam since these two polarization components lead to excitation to Zeeman levels with positive and negative $m_F$, respectively, with little overlap. Therefore, we expect their spectra to add up incoherently. More details can be found in Supplemental Material Sec.~I \cite{supplemental}.

\begin{figure}[!tbp]
\includegraphics[width=\linewidth]{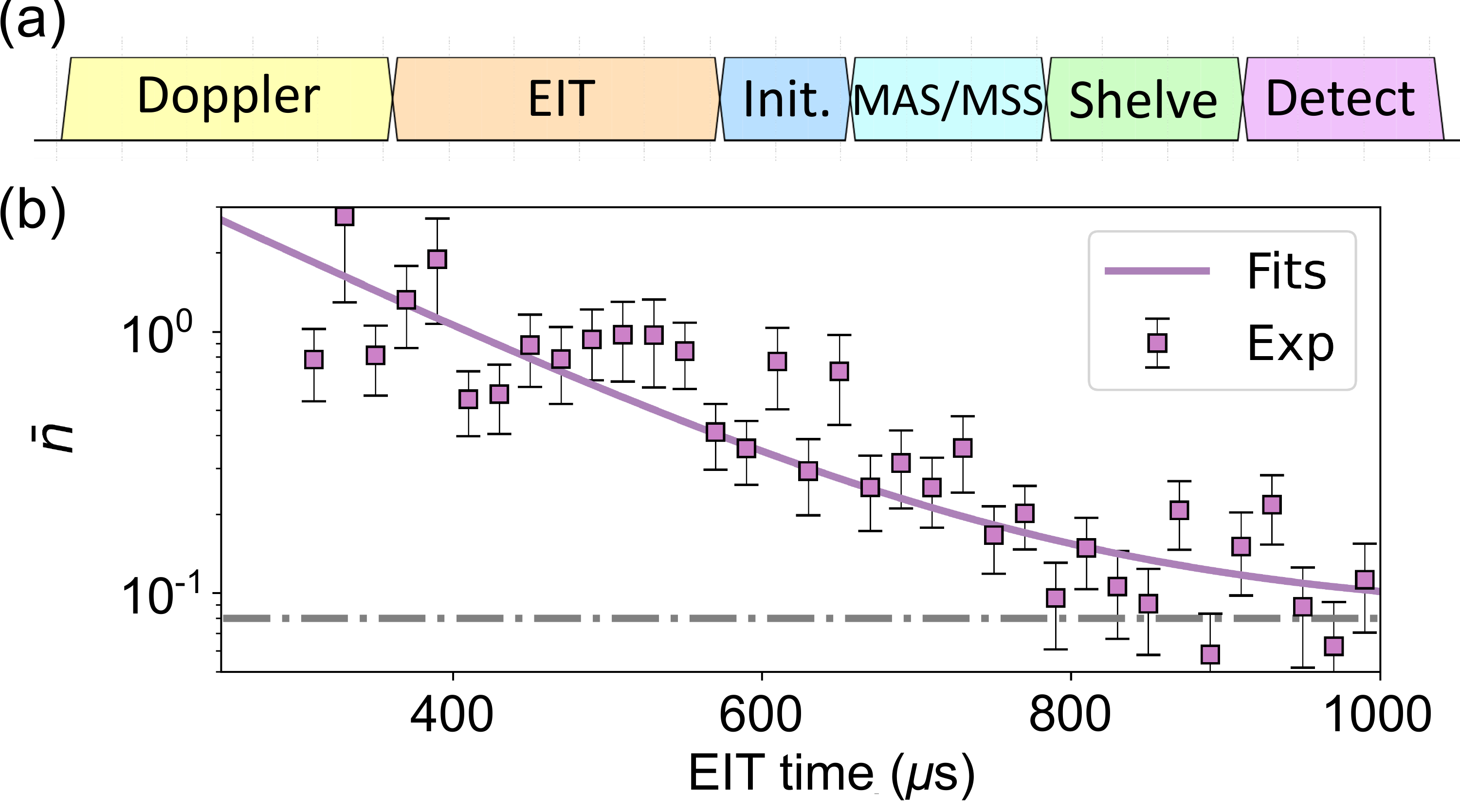}%
\caption{Single-ion EIT cooling. (a) EIT cooling pulse sequence. After Doppler and EIT cooling, we initialize the qubit state in $\ket{1}$ by optical pumping. Then we apply a motion-adding sideband (MAS) or motion-subtracting sideband (MSS) pulse before shelving detection. (b) EIT cooling dynamics with optimized parameters. We omit the data points for the first $300\,\mu$s since they are too high to be measured accurately by sideband ratio method.
We fit the data points to $\bar{n}(t) = (\bar{n}_{i}-\bar{n}_{f}) \exp(-t/\tau)+\bar{n}_{f}$, which yields a cooling limit $\bar{n}_{f}=0.08(2)$ (horizontal dashed line) and a $1/e$ cooling time $\tau=0.15(6)\,$ms. Error bars represent one standard deviation.}
\label{fig:coolingrate}
\end{figure}

\begin{figure*}[!tbp]
\includegraphics[width=0.8\linewidth]{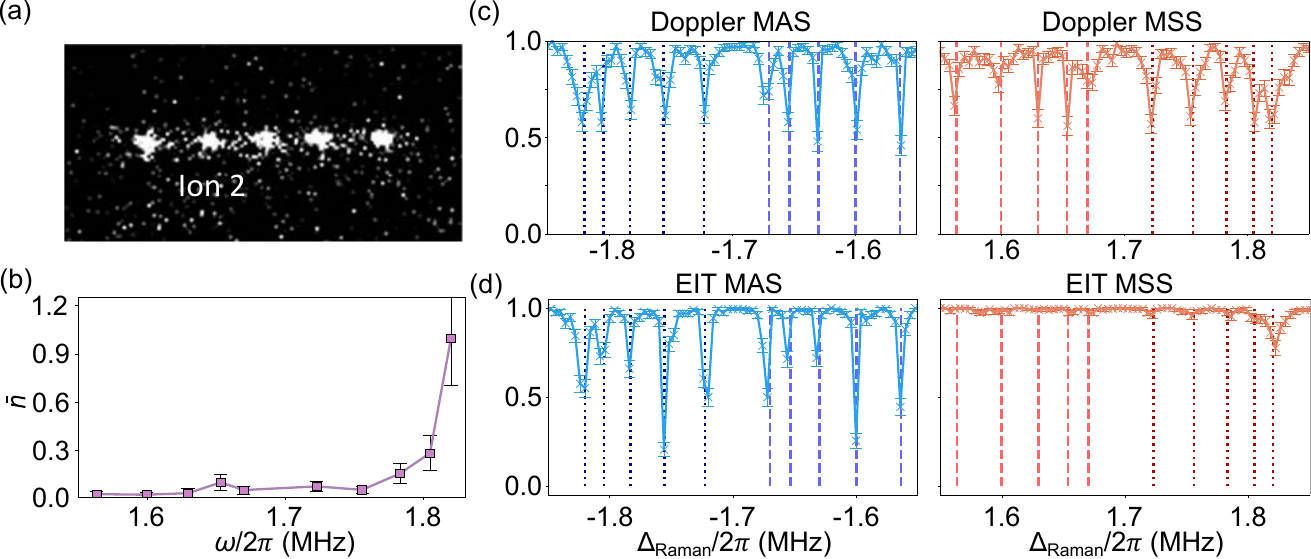}
\caption{Simultaneous EIT cooling of a multi-ion chain. (a) Image of a five-ion chain. Ion 2 is selectively addressed by Raman beams for average phonon number measurements. (b) Average phonon number $\bar{n}$ of all the ten radial modes calculated from MAS and MSS excitation probabilities shown in (d). Except for the $y$-center-of-mass mode (the highest frequency mode with the largest heating rate proportional to the ion number), the rest modes are cooled to an average $\bar{n}=0.09$. (c, d) MAS and MSS excitation spectra after Doppler cooling (c) and EIT cooling (d). Data points are connected by solid lines to guide the eye. Vertical dashed and dotted lines indicate the frequencies of the $x$ and $y$ modes, respectively. Error bars in (b-d) represent one standard deviation. }
\label{fig:5ion}
\end{figure*}

Figure~\ref{fig:Fanospec}(a) shows the combined spectrum, which is the total population of the $P_{1/2},F=2$ manifold as a function of $\Delta_{\mathrm{probe}}$. The calculation uses parameters close to experimental numbers, including $\Omega_{\mathrm{probe},+}=\Omega_{\mathrm{probe},-}=0.40\Gamma$, $\Omega_{\mathrm{pump},1}=1.5\Gamma$, $\Omega_{\mathrm{pump},2}=3.0\Gamma$, $\Delta_{\mathrm{pump}}=3.7\Gamma$. (See Supplemental Material Sec.~II for their calibrations.) We choose $\Delta=2\pi\times -2\,$MHz so that the two sets of bright resonant peaks merge together to allow efficient EIT cooling.
The Fano-like lineshape for the four possible $\Lambda$ systems is highlighted by the gray box and is further zoomed in by Fig.~\ref{fig:Fanospec}(b),
where the dark resonances are indicated by the vertical dashed lines and are labeled by the corresponding states.
We also experimentally verify this spectrum with a single $^{137}$Ba$^{+}$ ion by turning on the EIT cooling beams for $3\,$ms and count the scattered photon number which is proportional to the $P_{1/2}$ population. The lineshape of the experimental data represented by crossing marks agrees with the computed spectrum. Note that a nonzero background of $7.5$ photons has already been subtracted from the measured photon counts.

\emph{Cooling of a single ion---}
We perform EIT cooling with a single ion using the pulse sequence shown in Fig~\ref{fig:coolingrate}(a).
After Doppler cooling, the single ion is EIT cooled for a certain duration followed by an optical pumping pulse to reset the internal state to $\ket{1}$.
Then, we apply a motion-adding sideband (MAS) or motion-subtracting sideband (MSS) $\pi$ pulse before shelving detection.
We estimate $\bar{n}$ by using the ratio of the MAS and MSS excitation probabilities~\cite{RevModPhys.75.281}.

We set $\Delta_{\mathrm{probe}}=\Delta_{\mathrm{pump}}-\Delta_{\mathrm{Zeeman}}=3.45\Gamma$ to satisfy the dark resonance condition for the $\ket{1}$ and $\ket{a}$ states. To calibrate the optimal pump beam intensity for given transverse modes, we perform EIT cooling with varying pump beam power for a fixed duration of $2\,$ms, and measure $\bar{n}$ of the $x$ mode as a metric (see Supplemental Material Sec.~III \cite{supplemental}).
We determine the optimal values  $\Omega_{\mathrm{pump},1}=2.2\Gamma$ and $\Omega_{\mathrm{pump},2}=4.4\Gamma$. As for the probe intensity, in the absence of heating from environment, the cooling limit decreases monotonically as decreasing the probe beam intensity. In practice, external heating cannot be neglected and an optimal value of $\Omega_{\mathrm{probe}}=0.40\Gamma$ is determined from calibration.

Next, we investigate the cooling dynamics by measuring $\bar{n}$ for variable cooling time $t$ and show the experiment results in Fig.~\ref{fig:coolingrate}(b).
We fit the experimental data points to an exponential decay $\bar{n}(t) = (\bar{n}_{i}-\bar{n}_{f}) \exp(-t/\tau)+\bar{n}_{f}$, yielding $\tau=0.15(6)\,$ms, $\bar{n}_{i} = 11.8(2.6)$, and $\bar{n}_{f} = 0.08(2)$. At a longer duration of $2\,$ms, the $x$ mode phonon number is measured to be $\bar{n}= 0.08(5)$, consistent with the fitted $\bar{n}_{f}$.
The fitting result corresponds to a measured cooling rate $\gamma=7(2)\,$ms$^{-1}$, consistent with the theoretical calculation of $7.2\,$ms$^{-1}$ using Eq.~(\ref{eqa:gamma}).
As the $x$ and $y$ modes have similar frequencies, the EIT cooling optimized for the $x$ mode also cools the $y$ mode to an average phonon number of 0.15(7) for a $2\,$ms cooling duration. The higher $\bar{n}$ of the $y$ mode is primarily from its larger heating rate of $0.8(2)\,$ms$^{-1}$ compared to the $0.10(6)\,$ms$^{-1}$ of the $x$ mode.

\emph{Cooling of a multi-ion chain---}
Finally, we use the same parameters optimized for a single ion to EIT cool a five-ion chain, whose radial modes span a frequency range of about $270\,$kHz.
By using the same experimental sequence in Fig.~\ref{fig:coolingrate}(a), we simultaneously cool
all ten radial modes and show the final $\bar{n}$ in Fig.~\ref{fig:5ion}(b).
Raman pulses for sideband thermometry are only applied to ion 2 [labelled in Fig.~\ref{fig:5ion}(a)].
The MAS and MSS spectra after Doppler cooling and after EIT cooling are respectively shown in Fig.~\ref{fig:5ion}(c) and (d). The MSS excitation probabilities after EIT cooling are greatly suppressed for all the radial modes except for the $y$-center-of-mass mode due to its high heating rate proportional to the ion number. This mode is cooled to $\bar{n}=1.0(3)$, still significantly smaller than the Doppler limit of $\bar{n}\approx 10$.
The other nine modes are cooled to much lower $\bar{n}$ with an average number of 0.09.

In summary, we introduce an EIT cooling scheme applicable for atomic species possessing high nuclear spins at low fields. We successfully implement this scheme with both a single $^{137}\mathrm{Ba}^{+}$ ion and a five-ion chain, achieving average phonon numbers well below the Doppler cooling limit. Our approach can be extended to other atomic species possessing high nuclear spins and can provide efficient cooling for large ion crystals. More importantly, our scheme indicates that many ground states are not necessarily an issue for EIT cooling, but it offers freedoms to engineer the Fano-like spectrum to optimize cooling for specific scenarios. For example, we can perform simultaneous cooling of both radial and axial modes within different frequency ranges by adjusting the EIT beam propagation directions and tuning laser parameters. More details can be found in Supplemental Material Sec.~IV \cite{supplemental}.

\begin{acknowledgments}
During manuscript preparation, we became aware that the NIST Ion Storage Group is investigating EIT cooling with the $I=5/2$ $^{25}$Mg$^{+}$ isotope.

This work was supported by Innovation Program for Quantum Science and Technology (2021ZD0301601), Tsinghua University Initiative Scientific Research Program, and the Ministry of Education of China. L.M.D. acknowledges in addition support from the New Cornerstone Science Foundation through the New Cornerstone Investigator Program. Y.K.W. acknowledges in addition support from Tsinghua University Dushi program. P.Y.H. acknowledges the start-up fund from Tsinghua University.
\end{acknowledgments}
\providecommand{\noopsort}[1]{}\providecommand{\singleletter}[1]{#1}%

\end{document}


\title[EIT cooling]{Supplemental Material for ``Electromagnetically-Induced-Transparency \\ Cooling
of High-Nuclear-Spin Ions''}
\author{Chuanxin Huang}
\thanks{These authors contribute equally to this work}%
\affiliation{Center for Quantum Information, Institute for Interdisciplinary Information Sciences, Tsinghua University, Beijing 100084, P. R. China}
\author{Chenxi Wang}
\thanks{These authors contribute equally to this work}%
\affiliation{Center for Quantum Information, Institute for Interdisciplinary Information Sciences, Tsinghua University, Beijing 100084, P. R. China}
\author{Hongxuan Zhang}
\affiliation{Center for Quantum Information, Institute for Interdisciplinary Information Sciences, Tsinghua University, Beijing 100084, P. R. China}
\author{Hongyuan Hu}
\affiliation{Center for Quantum Information, Institute for Interdisciplinary Information Sciences, Tsinghua University, Beijing 100084, P. R. China}
\author{Zuqing Wang}
\affiliation{Center for Quantum Information, Institute for Interdisciplinary Information Sciences, Tsinghua University, Beijing 100084, P. R. China}
\author{Zhichao Mao}
\affiliation{HYQ Co., Ltd., Beijing 100176, P. R. China}
\author{Shijiao Li}
\affiliation{Center for Quantum Information, Institute for Interdisciplinary Information Sciences, Tsinghua University, Beijing 100084, P. R. China}
\author{Panyu Hou}
\affiliation{Center for Quantum Information, Institute for Interdisciplinary Information Sciences, Tsinghua University, Beijing 100084, P. R. China}
\affiliation{Hefei National Laboratory, Hefei 230088, P. R. China}
\author{Yukai Wu}
\affiliation{Center for Quantum Information, Institute for Interdisciplinary Information Sciences, Tsinghua University, Beijing 100084, P. R. China}
\affiliation{Hefei National Laboratory, Hefei 230088, P. R. China}
\author{Zichao Zhou}
\email{zichaozhou@mail.tsinghua.edu.cn}
\affiliation{Center for Quantum Information, Institute for Interdisciplinary Information Sciences, Tsinghua University, Beijing 100084, P. R. China}
\affiliation{Hefei National Laboratory, Hefei 230088, P. R. China}
\author{Luming Duan}
\email{lmduan@tsinghua.edu.cn}
\affiliation{Center for Quantum Information, Institute for Interdisciplinary Information Sciences, Tsinghua University, Beijing 100084, P. R. China}
\affiliation{Hefei National Laboratory, Hefei 230088, P. R. China}
\affiliation{New Cornerstone Science Laboratory, Beijing 100084, PR China}
\date{\today}

\maketitle

\section{Numerical simulation of excitation spectrum}
In general, to solve the quantum dynamics of multiple energy levels under multiple driving frequencies, we need to consider a master equation with a time-dependent Hamiltonian. As described in the main text, here for simplicity we approximate the steady state under the EIT pump and probe laser beams by considering the $\sigma_+$ and $\sigma_-$ polarization components of the probe beam separately. Since these two polarization components of the weak probe beam cause excitation to the positive and negative Zeeman levels, respectively, with little overlap, we expect their contribution to the excitation spectrum to add up incoherently. Below we present the result for the $\sigma_+$ polarization component, while that for the $\sigma_-$ component is symmetric.

\begin{figure}[h]
\includegraphics[width=0.45\textwidth]{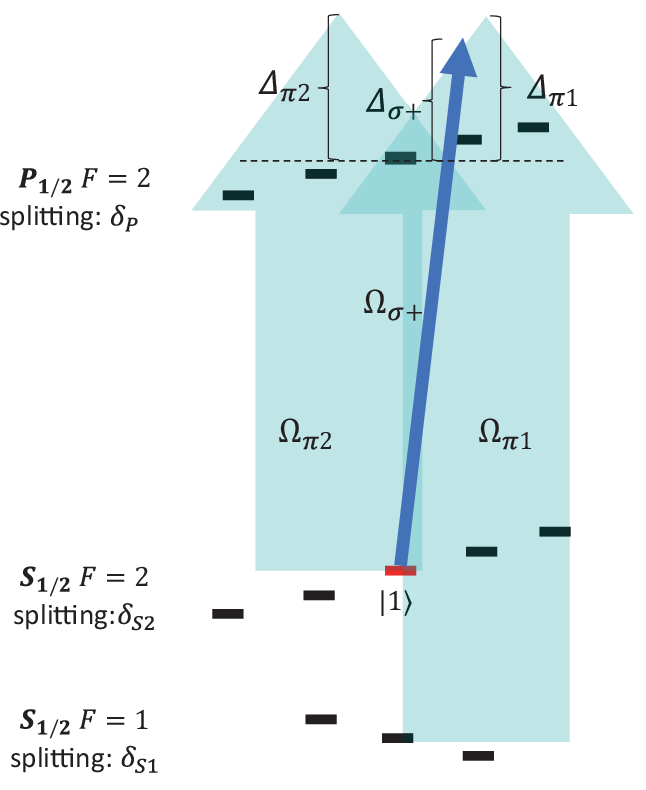}
\caption{Sketch of the energy levels and the laser components when the probe beam carries pure $\sigma_{+}$ polarization. The detuning is defined with respect to the $|S_{1/2},F=2,m_{F}=0\rangle\leftrightarrow|P_{1/2},F=2,m_{F}=0\rangle$ and the $|S_{1/2},F=1,m_{F}=0\rangle\leftrightarrow|P_{1/2},F=2,m_{F}=0\rangle$ transitions.}
\label{fig:MasterEquation}
\end{figure}

We consider all the 13 Zeeman levels in the $|S_{1/2},F=1\rangle$, $|S_{1/2},F=2\rangle$ and $|P_{1/2},F=2\rangle$ manifolds. As sketched in Fig.~\ref{fig:MasterEquation}, we have a $\pi$-polarized pump beam with two frequency components that couple $|S_{1/2},F=1\rangle$ and $|S_{1/2},F=2\rangle$, respectively, to $|P_{1/2},F=2\rangle$, with Rabi frequencies $\Omega_{\pi,1(2)}$, and we have a $\sigma_+$ polarization of the probe beam that couples $|S_{1/2},F=2\rangle$ to $|P_{1/2},F=2\rangle$ with a Rabi frequency $\Omega_{\sigma_+}$. Note that these three frequency components do not form any loop in the allowed transitions, therefore when moving into a suitable rotating frame, the Hamiltonian can be made time-independent.

Specifically, we can set the energy of $|P_{1/2},F=2,m_{F}=0\rangle$ to be the zero point
\begin{eqnarray}
\langle P_{1/2},F=2,m_{F}=0|H|P_{1/2},F=2,m_{F}=0\rangle = 0.
\end{eqnarray}
Then by requiring the Hamiltonian to be time-independent, we have
\begin{eqnarray}
\langle P_{1/2},F=2,m_{F}|H|P_{1/2},F=2,m_{F}\rangle = m_{F}\times (\Delta_{\pi, 2} - \Delta_{\sigma_+} + \delta_P), \\
\langle S_{1/2},F=2,m_{F}|H|S_{1/2},F=2,m_{F}\rangle = \Delta_{\pi, 2} + m_{F}\times(\Delta_{\pi, 2} - \Delta_{\sigma_+}+\delta_{S, 2}),\\
\langle S_{1/2},F=1,m_{F}|H|S_{1/2},F=1,m_{F}\rangle = \Delta_{\pi, 1} + m_{F}\times(\Delta_{\pi, 2} - \Delta_{\sigma_+}+\delta_{S, 1}),
\end{eqnarray}
where $\delta_{S,1} = -2\pi\times 4.7\,$MHz, $\delta_{S, 2} = 2\pi\times 4.7\,$MHz and $\delta_{P} = 2\pi\times 1.57\,$MHz are the Zeeman splitting.
The other coupling terms are given by
\begin{eqnarray}
\langle S_{1/2},F=2,m_{F}| H|P_{1/2},F=2,m_{F}\rangle = \frac{\Omega_{\pi 2}}{2} \mathrm{CG}(2,m_F;2,m_F), \\
\langle S_{1/2},F=1,m_{F}| H|P_{1/2},F=2,m_{F}\rangle = \frac{\Omega_{\pi 1}}{2} \mathrm{CG}(1,m_F;2,m_F), \\
\langle S_{1/2},F=2,m_{F}| H|P_{1/2},F=2,m_{F}+1\rangle = \frac{\Omega_{\sigma_{+}}}{2} \mathrm{CG}(2,m_F;2,m_F+1),
\end{eqnarray}
where $\mathrm{CG}(F_1,m_{F_1};F_2,m_{F_2})$ is the CG coefficient between $|S_{1/2},F_1,m_{F_1}\rangle$ and $|P_{1/2},F_2,m_{F_2}\rangle$.

The spontaneous emission of the $P_{1/2},F=2$ levels is modeled by Lindblad jump operators $L_{ij}=\sqrt{\gamma_{ij}}|i\rangle\langle j|$ where $i$ represents an $S_{1/2}$ state and $j$ represents a $P_{1/2},F=2$ state. The coefficient of each term is proportional to the corresponding CG coefficients between these levels such that the overall decay rate $\sum_i \gamma_{ij} = \Gamma=2\pi\times 14.6\,$MHz. Here we have taken into account the branching ratio of 0.268 from the $P_{1/2}$ levels to the irrelevant $D_{3/2}$ levels which are pumped back by a strong $650\,$nm beam.
Finally, we solve the steady state as the eigenstate of the Lindbladian with the eigenvalue of zero.

\section{Calibration of Rabi frequencies}

We calibrate the Rabi frequencies of the probe and the pump laser beams by their AC-stark shift on the $S_{1/2}$ levels, which can be measured from the Ramsey fringe.


Suppose the probe beam has all the polarization components with Rabi frequencies $\Omega_{\sigma_+}$, $\Omega_{\sigma_-}$ and $\Omega_{\pi}$. For a detuning $\Delta_{\mathrm{probe}}$ much larger than the Rabi frequency (with the additional CG coefficients) but much smaller than the hyperfine splitting, the differential AC Stark shift between the clock qubit levels $|S_{1/2},F=1,m_F = 0\rangle$ and $|S_{1/2},F=2,m_F = 0\rangle$
%
%
can be given by
\begin{equation}
\frac{\left| \mathrm{CG}(2,0;2,1) \Omega_{\sigma_+}\right|^2}{4(\Delta_{\mathrm{probe}}-\delta_{P})}
    +
\frac{\left| \mathrm{CG}(2,0;2,-1) \Omega_{\sigma_-}\right|^2}{4(\Delta_{\mathrm{probe}}+\delta_{P})}.
\end{equation}
Similarly, the differential AC Stark shift between the Zeeman qubit $|S_{1/2},F=1,m_F = 1 \rangle$ and $|S_{1/2},F=2,m_F = 2 \rangle$ is given by
\begin{equation}
\frac{\left| \mathrm{CG}(2,2;2,2) \Omega_{\pi}\right|^2}{4(\Delta_{\mathrm{probe}}+2\delta_{S,2}-2\delta_{P})}
    +
\frac{\left| \mathrm{CG}(2,2;2,1) \Omega_{\sigma_-}\right|^2}{4(\Delta_{\mathrm{probe}}+2\delta_{S,2}-\delta_{P})},
\end{equation}
and that between $|S_{1/2},F=1,m_F = -1 \rangle$ and $|S_{1/2},F=2,m_F = -2 \rangle$ is given by
\begin{equation}
\frac{\left| \mathrm{CG}(2,-2;2,-2) \Omega_{\pi}\right|^2}{4(\Delta_{\mathrm{probe}}-2\delta_{S,2}+2\delta_{P})}
    +
\frac{\left| \mathrm{CG}(2,-2;2,-1) \Omega_{\sigma_+}\right|^2}{4(\Delta_{\mathrm{probe}}-2\delta_{S,2}+\delta_{P})}.
\end{equation}

\begin{figure} [h]
		\includegraphics[scale=0.65]{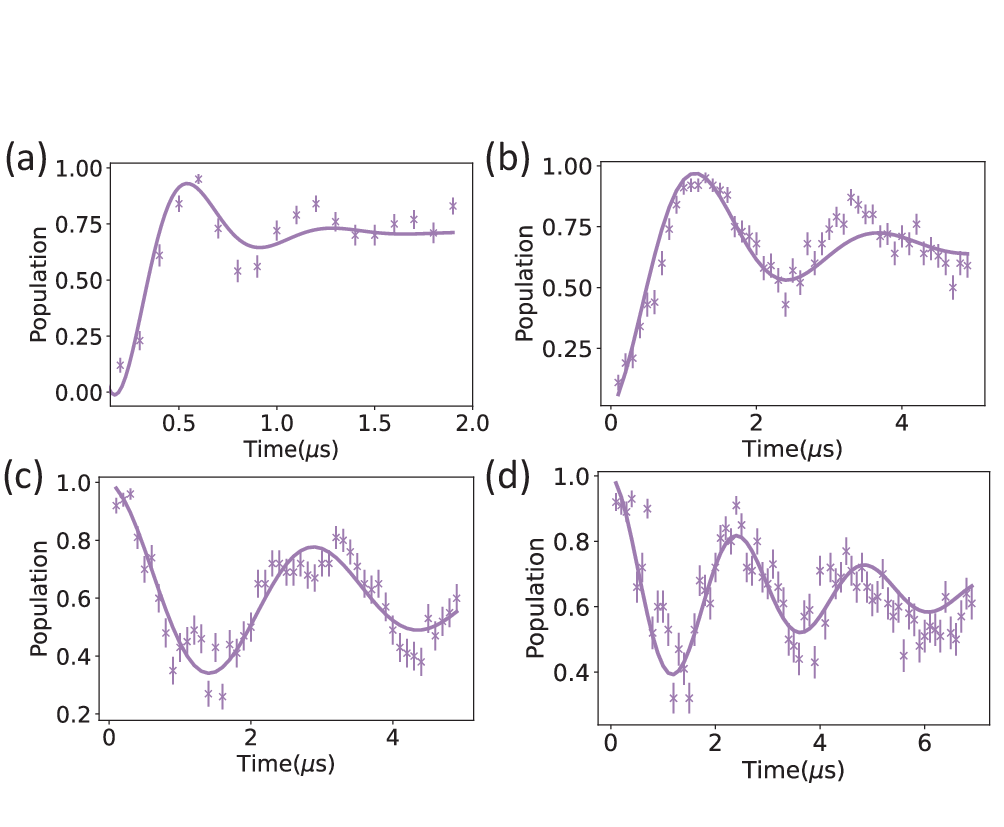}
	\caption{Fitting the AC Stark shift by a Ramsey experiment. (a) Ramsey fringes between $|S_{1/2},F=1,m_F = 0\rangle$ and $|S_{1/2},F=2,m_F = 0\rangle$ under the pump laser. (b-d) Ramsey fringes between $|S_{1/2},F=1,m_F = 0\rangle$ and $|S_{1/2},F=2,m_F = 0\rangle$, $|S_{1/2},F=1,m_F = 1 \rangle$ and $|S_{1/2},F=2,m_F = 2 \rangle$, and $|S_{1/2},F=1,m_F = -1 \rangle$ and $|S_{1/2},F=2,m_F = -2 \rangle$, respectively, under the probe laser.}
	\label{Fig_Ramsey}
\end{figure}

By measuring these frequency shifts on the clock or Zeeman qubits, we can determine $\Omega_{\sigma_+}$, $\Omega_{\sigma_-}$ and $\Omega_{\pi}$. In the experiment, we can set a microwave resonant to the clock or the Zeeman qubit when the probe laser is turned off. Then we turn on the probe laser and measure the Ramsey fringe as shown in Fig.~\ref{Fig_Ramsey}. The decrease in the oscillation amplitude comes from the dephasing of the qubit, the fluctuation of the probe laser intensity, and the spontaneous emission of the $P_{1/2}$ levels. We fit the Ramsey fringe by $A\sin^2 \delta t e^{-\gamma t} + B$, from which the frequency shift $\delta$ can be extracted.

Similarly, the Rabi frequency $\Omega_{\pi,1}$ of the $\pi$-polarized pump beam can be calibrated by its differential AC Stark shift on the clock qubit between $|S_{1/2},F=1,m_F = 0\rangle$ and $|S_{1/2},F=2,m_F = 0\rangle$
\begin{equation}
\frac{\left| \mathrm{CG}(1,0;2,0) \Omega_{\pi,1}\right|^2}{4\Delta_{\mathrm{pump,1}}}.
\end{equation}
The transition by $\Omega_{\pi,2}$ between $|S_{1/2},F=2,m_F = 0\rangle$ and $|P_{1/2},F=2,m_F = 0\rangle$ is forbidden by the selection rule, but by measuring the relative intensity of the two frequency components, we can obtain $\Omega_{\pi,1}\approx\Omega_{\pi,2}/2$.

\section{Optimization of cooling parameters}
\begin{figure}[h]
\includegraphics[width=0.65\textwidth]{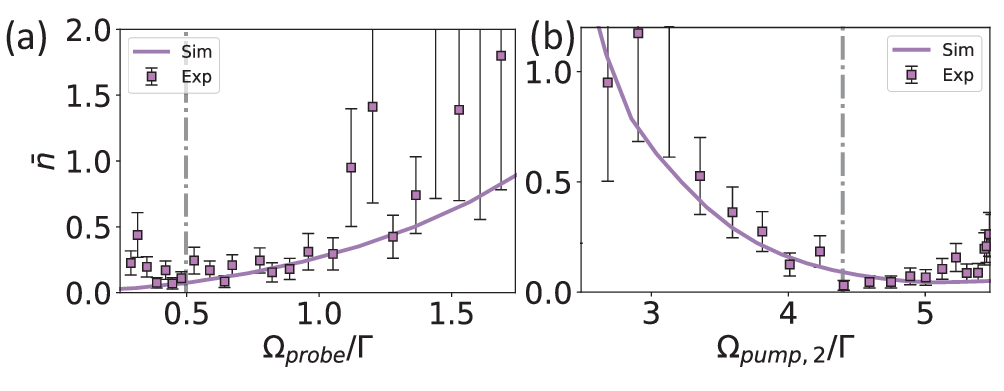}
\caption{We scan the intensity of (a) the probe beam and (b) the pump beam to find the optimal cooling parameters in terms of the final phonon number after $2\,$ms EIT cooling (squares). Solid curves represent the numerical results without considering external heating sources.}
\label{fig:OptiAmp}
\end{figure}

In Fig.~\ref{fig:OptiAmp} we scan the intensity of the probe and the pump laser for the minimal final phonon number after $2\,$ms EIT cooling.
Ideally without additional heating sources, the theoretical cooling limit decreases monotonically as the probe beam weakens. However, in practice due to the competition with the external heating sources, $\Omega_{\mathrm{probe}}=0.40\Gamma$ is chosen for the optimal cooling limit as shown in Fig.~\ref{fig:OptiAmp}(a).

As for the pump beam, theoretically for an ideal $\Lambda$ system, the optimal pumping intensity is to have the bright and the dark resonance separated by the phonon frequency. Here with the existence of multiple bright resonant peaks with different AC Stark shifts, it is more difficult to determine the optimal pump intensity analytically, so instead we scan it experimentally and obtain an optimal Rabi rate of about $\Omega_{\mathrm{pump},2}=4.4\Gamma$. Also it would be desirable to set the Rabi rates of the two frequency components of the pump beam to about the same, such that the two sets of Fano-like spectrum can have comparable heights and we can adjust them to overlap with each other for the most efficient EIT cooling. However, in this experiment our choice of $\Omega_{\mathrm{pump},2}/\Omega_{\mathrm{pump},1}\approx 2$ is due to our electro-optic modulator (EOM) that generates these two frequency components. Specifically, we use the 0th and the 2nd sidebands of the EOM for these two components and their ratio is restricted by the RF power on the EOM. To further enhance the 2nd EOM sideband and to balance the two Rabi frequencies, higher RF power is needed and can induce additional heating and potentially decrease the stability of our laser intensity.

\section{Engineering Fano-like spectrum for simultaneous cooling of radial and axial modes}
\begin{figure}[h]
\includegraphics[width=0.5\textwidth]{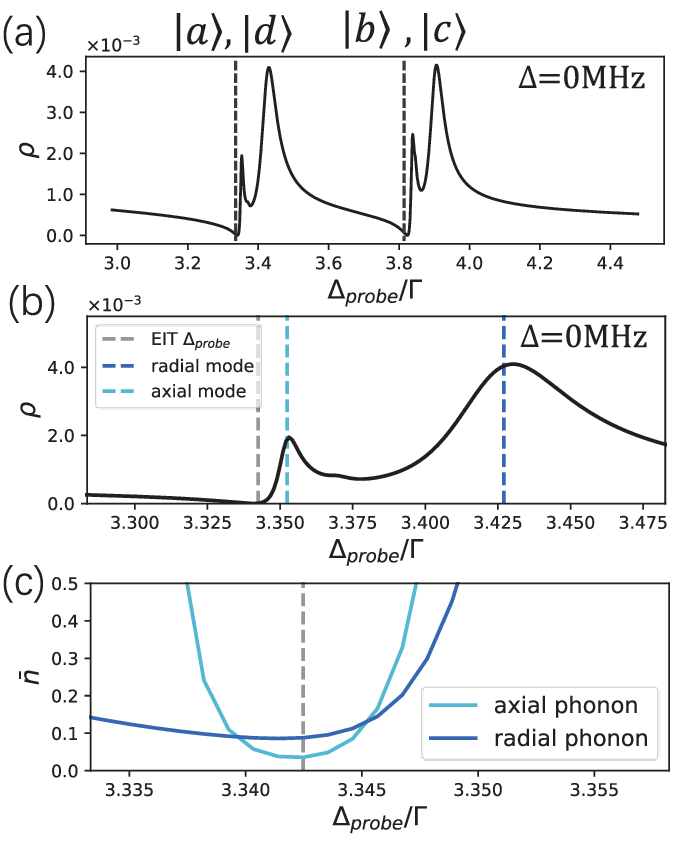}
\caption{(a) Theoretical spectrum at $\Omega_{\mathrm{probe},+}=\Omega_{\mathrm{probe},-}=0.15\Gamma$, $\Omega_{\mathrm{pump},1}=2.2\Gamma$, $\Omega_{\mathrm{pump},2}=1.5\Gamma$, $\Delta_{\mathrm{pump}}=3.6\Gamma$ and $\Delta=0$. (b) A zoom-in for the spectrum around the bright resonant peaks of $|a\rangle$ and $|d\rangle$. (c) Cooling limit for the axial mode and the radial mode vs. the probe detuning. The sensitivity for the axial mode is higher since it has lower frequency and corresponds to a narrower bright resonant peak.}
\label{fig:FourPeak}
\end{figure}

In the main text, we choose $\Delta=-2\pi\times 2\,$MHz between the two frequency components of the pump beam such that the two sets of bright resonant peaks roughly coincide with each other. For general $\Delta$, we will see all the four bright resonant peaks as shown in Fig.~\ref{fig:FourPeak}(a). While this type of spectra may not be optimal for EIT cooling of a given phonon frequency, here we mention that it may be advantageous for the simultaneous cooling of two frequency bands, e.g. the radial and the axial modes, which may be desired for the future large-scale ion trap quantum computing and simulation.
Specifically, if we engineer the spectrum such that the two bright resonant peaks are centered around the desired two phonon bands as shown in Fig.~\ref{fig:FourPeak}(b), e.g. in our case an axial mode at $2\pi\times 0.2\,$MHz and a radial mode at $2\pi\times 1.7\,$MHz, simultaneous EIT cooling of both modes can be achieved with a theoretical phonon number of $n_a=0.035$ and $n_r = 0.086$.

Note that some previous works with $I=0$ \cite{PhysRevA.98.023424} or $I=1/2$ \cite{PhysRevLett.125.053001,PhysRevLett.126.023604} also allow to engineer the Fano-like spectrum via two possible $\Lambda$ systems. However, a major difference is that these systems usually posses natural frequency or polarization selection such that the pump beam for the additional $\Lambda$ system does not destroy the dark state of the original $\Lambda$ system. For example, in Ref.~\cite{PhysRevA.98.023424} for ${}^{40}\mathrm{Ca}^+$ ions with $I=0$, the two $\Lambda$ systems are formed by $\ket{S_{1/2},m_F=1/2}$, $\ket{S_{1/2},m_F=-1/2}$ and $\ket{P_{1/2},m_F=1/2}$, and $\ket{S_{1/2},m_F=-1/2}$, $\ket{D_{3/2},m_F=1/2}$ and $\ket{P_{1/2},m_F=-1/2}$, respectively. The $866\,$nm pump beam for the latter, coupling $D_{3/2}$ to $P_{1/2}$, does not affect the dark state of the former in the $S_{1/2}$ states because of the large frequency detuning. Similarly, in Refs.~\cite{PhysRevLett.125.053001,PhysRevLett.126.023604}  for ${}^{171}\mathrm{Yb}^+$ ions with $I=1/2$ the two $\Lambda$ systems are formed by $\ket{S_{1/2},F=1,m_F=0}$, $\ket{S_{1/2},F=1,m_F=1}$ and $\ket{P_{1/2},F=0,m_F=0}$, and $\ket{S_{1/2},F=1,m_F=0}$, $\ket{S_{1/2},F=1,m_F=-1}$ and $\ket{P_{1/2},F=0,m_F=0}$, respectively. The $\sigma_\pm$ pump beams for these two $\Lambda$ systems do not affect the dark state of each other because of the polarization selection rule. In contrast, in our work with $I=3/2$, our multiple $\Lambda$ systems share lower states within the same hyperfine structure manifold, so that a simple frequency selection cannot be achieved. Also, due to the existence of multiple Zeeman levels for the upper states, a simple polarization selection cannot be obtained, either. In this sense, our main contribution is to design the laser configurations for EIT cooling of high-nuclear-spin ions as described in the main text, while the capability of engineering the Fano-like spectrum for simultaneous cooling of multiple phonon modes is more of a side effect.
\section{Generalization to higher nuclear spins}
Our scheme can be readily applied to ion species with higher nuclear spins $I=5/2,\,7/2,\,9/2,\,\cdots$. Still we use $\pi$-polarized EIT pump laser with two frequency components to couple $S_{1/2},F=I\pm 1/2$ levels to $P_{1/2},F=I+1/2$. Due to the dipole-forbidden transition between $\ket{S_{1/2},F=I+1/2,m_F=0}$ and $\ket{P_{1/2},F=I+1/2,m_F=0}$, the dark state will again be $\ket{S_{1/2},F=I+1/2,m_F=0}$. Then under the same $\sigma_{\pm}$-polarized EIT probe beam, the same four possible $\Lambda$ systems can be formed as shown in Fig.~1(c) of the main text and a similar Fano-like spectrum can be obtained for efficient EIT cooling.

\providecommand{\noopsort}[1]{}\providecommand{\singleletter}[1]{#1}%